\documentclass[prd,floats,aps,twocolumn,epsf,graphicx]{revtex4}
\usepackage{epsfig}
\usepackage{graphicx}

\begin{document}

\title{The fastest way to circle a black hole}
\author{Shahar Hod}
\address{The Ruppin Academic Center, Emeq Hefer 40250, Israel}
\address{}
\address{The Hadassah Institute, Jerusalem 91010, Israel}
\date{\today}

\begin{abstract}

\ \ \ Black-hole spacetimes with a ``photonsphere", a hypersurface
on which massless particles can orbit the black hole on circular
null geodesics, are studied. We prove that among all possible
trajectories (both geodesic and non-geodesic) which circle the
central black hole, the null circular geodesic is characterized by
the {\it shortest} possible orbital period as measured by asymptotic
observers. Thus, null circular geodesics provide the fastest way to
circle black holes. In addition, we conjecture the existence of a
universal lower bound for orbital periods around compact objects (as
measured by flat-space asymptotic observers): $T_{\infty}\geq 4\pi
M$, where $M$ is the mass of the central object. This bound is
saturated by the null circular geodesic of the maximally rotating
Kerr black hole.
\end{abstract}
\bigskip
\maketitle

\section{Introduction}

The motion of test particles in black-hole spacetimes has been
extensively studied for more than four decades, see
\cite{Bar,Chan,Shap,CarC} are references therein. Of particular
importance are geodesic motions which provide valuable information
on the structure of the spacetime geometry. Circular null geodesics
(also known as ``photonspheres") are especially interesting from
both an astrophysical and theoretical points of view
\cite{Notephoton}. As pointed out in \cite{CarC}, the optical
appearance to external observers of a star undergoing gravitational
collapse is related to the properties of the unstable circular null
geodesic \cite{CarC,Pod,Ame}.

Furthermore, null circular geodesics are closely related to the
characteristic oscillation modes of black holes (see e.g.
\cite{Noll,BerCar} for detailed reviews). These quasinormal
resonances can be interpreted in terms of null particles trapped at
the unstable circular orbit and slowly leaking out
\cite{CarC,Mash,Goeb,Hod1,Dol1,Dec}. The real part of the complex
quasinormal frequencies is related to the angular velocity at the
unstable null geodesic (as measured by asymptotic observers) while
the imaginary part of the resonances is related to the instability
timescale of the orbit \cite{CarC,Mash,Goeb,Hod1,Dol1,Dec} (or the
inverse Lyapunov exponent of the geodesic \cite{CarC}).

An important physical quantity for the analysis of circular orbits
in black-hole spacetimes is the angular frequency $\Omega_{\infty}$
of the orbit as measured by asymptotic observers. In this paper we
shall reveal an interesting property of null circular geodesics
which is related to this important quantity: We shall show that null
circular geodesics provide the fastest way to circle black holes.
More explicitly, we shall prove that the null circular geodesic of a
black-hole spacetime is characterized by the {\it shortest} possible
orbital period (the largest orbital frequency) as measured by
asymptotic observers.

It is worth pointing out that the orbital period $T$ around a
spherical compact object must be bounded from below by the mass $M$
of the central object: Suppose the compact object has radius $R$,
then obviously $T\geq 2\pi R$. (We shall use natural units in which
$G=c=1$). In addition, the central object must be larger than its
gravitational radius, $R\geq 2M$. Thus, the orbital period must be
bounded from below by
\begin{equation}\label{Eq1}
T\geq 4\pi M\  .
\end{equation}

However, it should be realized that the above reasoning is actually
too naive -- it does not take into account the possible influence of
the spacetime curvature (in the region near the surface of the
compact object) on the orbital period. Due to the influence of the
gravitational time dilation effect (redshift), the orbital period
$T_{\infty}$ as measured by asymptotic observers would actually be
{\it larger} than $2\pi R$. Moreover, we shall show below that due
to the influence of the redshift factor, the circular orbit with the
shortest orbital period (as measured by asymptotic observers) is
{\it distinct} from the circular orbit with the smallest
circumference (that is, $r_{\text{fast}}\neq R$ in general, where
$r_{\text{fast}}$ to be determined below is the radius of the
circular trajectory with the shortest orbital period).

\section{Spherically symmetric spacetimes}

We shall first consider static spherically symmetric asymptotically
flat black-hole spacetimes. The line element may take the following
form in Schwarzschild coordinates \cite{CarC,Hodm}
\begin{equation}\label{Eq2}
ds^2=-f(r) dt^2 +{{1}\over{g(r)}}dr^2+r^2(d\theta^2 +\sin^2\theta
d\phi^2)\ ,
\end{equation}
where the metric functions $f(r)$ and $g(r)$ depend only on the
Schwarzschild areal coordinate $r$. These functions should be
determined by solving the field equations. Since we do not specify
the field equations, our results would be valid for all spherically
symmetric asymptotically flat black holes. We note, in particular,
that we do {\it not} assume $g(r)=f(r)$ (a property which
characterizes the familiar Schwarzschild and Reissner-Nordstr\"om
spacetimes) and thus our results would be applicable to hairy
black-hole configurations as well [in these spacetimes $g(r)\neq
f(r)$, see \cite{BizCol,Lavr,Green,Stra,Volkov} and references
therein]. Asymptotic flatness requires that as $r \to \infty$,
\begin{equation}\label{Eq3}
f(r) \to 1\ \ \ {\text{and}}\ \ \ g(r) \to 1\  ,
\end{equation}
and a regular event horizon at $r=r_H$ requires \cite{Nun}
\begin{equation}\label{Eq4}
f(r_H)=g(r_H)=0\  .
\end{equation}

We shall now consider the following question: What is the radius
$r=r_{\text{fast}}$ of the circular trajectory \cite{Notegeonon}
around a central black hole which has the shortest orbital period
(the largest orbital frequency) as measured by asymptotic observers?
It is obvious that in order to minimize the orbital period for a
given radius $r$, one should move as close as possible to the speed
of light \cite{Notenongeo}. In this case, the orbital period as
measured by asymptotic observers can be obtained from Eq.
(\ref{Eq2}) with $ds=dr=d\theta=0$ and $\Delta\phi=2\pi$:
\begin{equation}\label{Eq5}
T(r)\equiv T_{\infty}(r)={{2\pi r}\over{\sqrt{f(r)}}}\  ,
\end{equation}
where $r$ is the radius of the circular trajectory. The denominator
of (\ref{Eq5}) represents the well-known redshift factor.

Note that for asymptotically flat spacetimes $T(r)\to\infty$ at both
$r\to r_H$ and $r\to\infty$. Thus, the function $T(r)$ must have a
minimum for some finite $r=r_{\text{fast}}>r_H$. [In particular, the
circular trajectory with the smallest possible radius ($r\to r_H$)
is characterized by an infinite orbital period as measured by
asymptotic observers. Thus, the flat-space reasoning which leads to
$T_{\text{min}}=\text{min}_r\{2\pi r\}=2\pi R$ for a compact object
of radius $R$ is certainly not correct for the curved region near a
black hole.]

The circular trajectory around the central black hole which has the
shortest orbital period is characterized by
\begin{equation}\label{Eq6}
T'(r=r_{\text{fast}})=0\  ,
\end{equation}
where a prime denotes a derivative with respect to areal radius $r$.
This yields the characteristic equation
\begin{equation}\label{Eq7}
r_{\text{fast}}={{2f(r_{\text{fast}})}\over{f'(r_{\text{fast}})}}\
\end{equation}
for the circular trajectory with the shortest orbital period.

We shall now prove that it is the null circular geodesic of the
black-hole spacetime which has the shortest orbital period as
measured by asymptotic observers. To that end, we shall follow the
analysis of \cite{Chan,Shap,CarC} and compute the location
$r=r_{\gamma}$ of the null circular geodesic for a black-hole
spacetime described by the line element (\ref{Eq2}). The Lagrangian
describing the geodesics in the spacetime (\ref{Eq2}) is given by
\begin{equation}\label{Eq8}
2{\cal L}=-f(r)\dot t^2+{{1}\over{g(r)}}\dot r^2+r^2\dot\phi^2\ ,
\end{equation}
where a dot denotes a derivative with respect to the affine
parameter along the geodesic. The generalized momenta derived from
this Lagrangian are given by \cite{Chan,Shap,CarC}
\begin{equation}\label{Eq9}
p_t=-f(r)\dot t\equiv -E={\text{const}}\  ,
\end{equation}
\begin{equation}\label{Eq10}
p_{\phi}=r^2\dot\phi\equiv L={\text{const}}\  ,
\end{equation}
and
\begin{equation}\label{Eq11}
p_r={{1}\over{g(r)}}\dot r\  .
\end{equation}
The Lagrangian is independent of both $t$ and $\phi$. This implies
that $E$ and $L$ are constants of the motion. The Hamiltonian of the
system is given by \cite{Chan,Shap,CarC} ${\cal H}=p_t\dot t
+p_r\dot r +p_{\phi}\dot\phi-{\cal L}$, which implies
\begin{eqnarray}\label{Eq12}
2{\cal H}=-E\dot t+L\dot\phi+{{1}\over{g(r)}}\dot
r^2=\delta={\text{const}}\ ,
\end{eqnarray}
where $\delta=0$ for null geodesics and $\delta=1$ for timelike
geodesics. Substituting Eqs. (\ref{Eq9})-(\ref{Eq11}) into
(\ref{Eq12}), one finds
\begin{equation}\label{Eq13}
\dot r^2={{g(r)}\over{f(r)}}E^2\Big[1-b^2{{f(r)}\over{r^2}}\Big]\
\end{equation}
for null geodesics, where $b\equiv L/E=\text{const}$.

Circular geodesics are characterized by $\dot r^2=(\dot r^2)'=0$
\cite{Chan,Shap,CarC}. The requirement $(\dot r^2)'=0$ yields the
equation
\begin{equation}\label{Eq14}
r_{\gamma}={{2f(r_{\gamma})}\over{f'(r_{\gamma})}}\
\end{equation}
for the null circular geodesic. Taking cognizance of the fact that
Eq. (\ref{Eq14}) for the null circular geodesic is {\it identical}
to Eq. (\ref{Eq7}) for the fastest circular trajectory (the circular
trajectory with the shortest orbital period), one realizes that the
null circular geodesic is characterized by the shortest possible
orbital period as measured by asymptotic observers:
\begin{equation}\label{Eq15}
r_{\text{fast}}=r_{\gamma}\  .
\end{equation}

It is worth noting that Eq. (\ref{Eq6}) [and thus also Eq.
(\ref{Eq7})] may have several solutions. Nevertheless, our
conclusion that the circular trajectory around the black hole with
the shortest orbital period coincides with a null circular geodesic
of the black-hole spacetime still holds true in such cases as well:
The fact that Eq. (\ref{Eq14}) for the null circular geodesic(s) is
identical to Eq. (\ref{Eq7}) for the circular trajectory(ies) with
$T'(r)=0$, implies that if $T(r)$ has several extremum points [that
is, Eq. (\ref{Eq7}) has several solutions], then the black-hole
spacetime would be characterized by several circular null geodesics
[that is, Eq. (\ref{Eq14}) would also have (the same) several
solutions]. The main point is that {\it any} solution of Eq.
(\ref{Eq7}) is {\it also} a solution of the identical equation
(\ref{Eq14}). Thus, the circular trajectory with the global minimum
of $T(r)$ [this trajectory is characterized by the relation
$rf'-2f=0$, see Eq. (\ref{Eq7})] must coincide with (one of) the
null circular geodesic(s) of the black-hole spacetime [these null
circular geodesics are {\it also} characterized by the {\it same}
relation, $rf'-2f=0$, see Eq. (\ref{Eq14})].

The shortest possible orbital period around the black hole as
measured by asymptotic observers is obtained by substituting the
optimal radius (\ref{Eq15}) back into Eq. (\ref{Eq5}). It can be
shown \cite{Hodto} that all spherical black holes (including hairy
solutions \cite{BizCol,Lavr,Green,Stra,Volkov}) satisfy the
inequality
\begin{equation}\label{Eq16}
f(r_{\gamma})\leq {1\over 3}\  .
\end{equation}
[The case $f(r_{\gamma})=1/3$ corresponds to the (``bare")
Schwarzschild black hole.] Taking cognizance of Eqs. (\ref{Eq5}),
(\ref{Eq15}), and (\ref{Eq16}), one finds that the shortest possible
orbital period around a spherical black hole must conform to the
lower bound
\begin{equation}\label{Eq17}
T_{\text{min}}\equiv\text{min}_r\{T(r)\}\geq 2\sqrt{3}\pi
r_{\gamma}\ .
\end{equation}
[Compare this with the {\it weaker} flat-space bound $T\geq 2\pi R$
discussed above.]

\section{Rotating Kerr black holes}

Our conclusions can be generalized to include non-spherically
symmetric black-hole spacetimes. We shall now analyze circular
trajectories around rotating Kerr black holes. In Boyer-Lindquist
coordinates the line element of the Kerr spacetime takes the form
\cite{Chan,Kerr}
\begin{eqnarray}\label{Eq18}
ds^2=-\Big(1-{{2Mr}\over{\rho^2}}\Big)dt^2-{{4Mar\sin^2\theta}\over{\rho^2}}dt
d\phi+{{\rho^2}\over{\Delta}}dr^2 \nonumber \\
+\rho^2d\theta^2+\Big(r^2+a^2+{{2Ma^2r\sin^2\theta}\over{\rho^2}}\Big)\sin^2\theta
d\phi^2,
\end{eqnarray}
where $\Delta\equiv r^2-2Mr+a^2$ and $\rho^2\equiv
r^2+a^2\cos^2\theta$. Here $M$ and $a$ are the black-hole mass and
angular momentum per unit mass, respectively. The event (outer)
horizon of the black hole is located at $r_H=M+(M^2-a^2)^{1/2}$ and
the stationary limit surface of the exterior spacetime is located at
$r_S=M+(M^2-a^2\cos^2\theta)^{1/2}$.

We shall consider circular orbits in the equatorial plane of the
black hole. These are characterized by $\theta=\pi/2$, which implies
$\rho=r$ and $r_S=2M$. Again, it is obvious that in order to
minimize the orbital period for a given radius $r$, one should move
as close as possible to the speed of light \cite{Notenongeo}. In
this case, the orbital period as measured by asymptotic observers
can be obtained from Eq. (\ref{Eq18}) with $ds=dr=d\theta=0$ and
$\Delta\phi=\pm2\pi$ \cite{Notephi}:
\begin{equation}\label{Eq19}
T(r)=2\pi{{\sqrt{\Delta}\mp{{2Ma}\over{r}}}\over{1-{{2M}\over{r}}}}\
,
\end{equation}
where the upper/lower signs correspond to
co-rotating/counter-rotating orbits. [Note that $T(r\to
2M)\to\infty$ for a counter-rotating orbit as the stationary surface
$r=2M$ is approached.]

The circular trajectory around the central black hole which has the
shortest orbital period is characterized by
\begin{equation}\label{Eq20}
T'(r=r_{\text{fast}})=0\  ,
\end{equation}
which yields the equation
\begin{equation}\label{Eq21}
r(r-2M)(r-3M)\pm2Ma(\sqrt\Delta\mp a)=0\  .
\end{equation}
It is straightforward to show that a solution of the condition
(\ref{Eq21}) is given by
\begin{equation}\label{Eq22}
r_{\text{fast}_{\pm}}=2M[1+\cos[{2\over 3}\cos^{-1}(\mp{a/M})]]\ .
\end{equation}

We shall now prove that the null circular geodesic of the Kerr black
hole is characterized by the shortest orbital period as measured by
asymptotic observers. To that end, we shall follow the analysis of
\cite{Chan,Shap,CarC} and compute the location $r=r_{\gamma}$ of the
null circular geodesic of the Kerr spacetime. The Lagrangian
describing the geodesics in the spacetime (\ref{Eq18}) is given by
\begin{equation}\label{Eq23}
2{\cal L}=g_{tt}\dot t^2+2g_{t\phi}\dot t \dot\phi +g_{rr}\dot
r^2+g_{\phi\phi}\dot\phi^2\ .
\end{equation}
The generalized momenta derived from this Lagrangian are given by
\cite{Chan,Shap,CarC}
\begin{equation}\label{Eq24}
p_t=g_{tt}\dot t+g_{t\phi}\dot\phi \equiv -E={\text{const}}\  ,
\end{equation}
\begin{equation}\label{Eq25}
p_{\phi}=g_{t\phi}\dot t+ g_{\phi\phi}\dot\phi\equiv
L={\text{const}}\ ,
\end{equation}
and
\begin{equation}\label{Eq26}
p_r=g_{rr}\dot r\  .
\end{equation}
The Lagrangian is independent of both $t$ and $\phi$. This implies
that $E$ and $L$ are constants of the motion. The Hamiltonian of the
system is given by \cite{Chan,Shap,CarC} ${\cal H}=p_t\dot t
+p_r\dot r +p_{\phi}\dot\phi-{\cal L}$, which implies
\begin{eqnarray}\label{Eq27}
2{\cal H}=-E\dot t+L\dot\phi+{{r^2}\over{\Delta}}\dot
r^2=\delta={\text{const}}\ ,
\end{eqnarray}
where $\delta=0$ for null geodesics and $\delta=1$ for timelike
geodesics. Substituting Eqs. (\ref{Eq24})-(\ref{Eq26}) into
(\ref{Eq27}), one finds
\begin{equation}\label{Eq28}
\dot
r^2=E^2\Big[1+{{a^2-b^2}\over{r^2}}+{{2M(a-b)^2}\over{r^3}}\Big]\
\end{equation}
for null geodesics.

The requirement $\dot r^2=0$ for a null circular geodesic
\cite{Chan,Shap,CarC} yields
\begin{equation}\label{Eq29}
b_{\pm}={{\sqrt{\Delta}\mp{{2Ma}\over{r}}}\over{1-{{2M}\over{r}}}}\
,
\end{equation}
which in view of Eq. (\ref{Eq19}) implies
\begin{equation}\label{Eq30}
b_{\pm}={{T(r_{\gamma})}\over{2\pi}}=\Omega^{-1}(r_{\gamma})\  ,
\end{equation}
where $\Omega\equiv\Omega_{\infty}$ is the angular frequency of the
orbit as measured by asymptotic observers. The requirement $(\dot
r^2)'=0$ \cite{Chan,Shap,CarC} yields the equation
\begin{equation}\label{Eq31}
r^2-3Mr\pm2a\sqrt{Mr}=0\  .
\end{equation}
It is straightforward to show that a solution of this equation is
given by \cite{Chan,Shap,Mash}:
\begin{equation}\label{Eq32}
r_{\gamma_{\pm}}=2M[1+\cos[{2\over 3}\cos^{-1}(\mp{a/M})]]\ .
\end{equation}
Taking cognizance of Eq. (\ref{Eq22}) for the radius
$r_{\text{fast}}$ of the fastest circular trajectory, one realizes
that the null circular geodesic (\ref{Eq32}) around a Kerr black
hole is characterized by the shortest possible orbital period as
measured by asymptotic observers:
\begin{equation}\label{Eq33}
r_{\text{fast}}=r_{\gamma}\  .
\end{equation}

The shortest possible orbital period around a Kerr black hole is
obtained by substituting the optimal radius (\ref{Eq22}) back into
Eq. (\ref{Eq19}). One then finds that
$T_{\text{min}}\equiv\text{min}_r\{T(r;a/M)\}$ monotonically
decreases as the ratio $a/M$ increases: The spherically symmetric
Schwarzschild black hole is characterized by
\begin{equation}\label{Eq34}
T_{\text{min}}=6\sqrt{3}\pi M\
\end{equation}
[compare this with the naive (and weaker) bound (\ref{Eq1}) for
spherical objects], while the maximally rotating Kerr black hole is
characterized by
\begin{equation}\label{Eq35}
T_{\text{min}}=4\pi M\  .
\end{equation}



\section{Summary}

In summary, we have studied circular orbits around central black
holes. In particular, we have analyzed the dependence of the orbital
period on the radius of the circular trajectory. It was shown that
the {\it null} circular geodesic is characterized by the {\it
shortest} possible orbital period as measured by asymptotic
observers. We therefore conclude that null circular geodesics
provide the fastest way to circle black holes.

It was pointed out in \cite{CarC} that intriguing physical phenomena
could occur in curved spacetimes for which there is a timelike
circular geodesic with an angular frequency which equals the angular
frequency of the unstable null circular geodesic. For instance, this
would raise the interesting possibility of exciting the black-hole
quasinormal frequencies by orbiting particles, possibly leading to
instabilities of the spacetime \cite{CarC}. However, our results
rule out this scenario for black-hole spacetimes with only one null
circular geodesic. In particular, we have shown that the null
circular geodesic is characterized by the largest angular frequency
as measured by asymptotic observers -- no other circular trajectory
(be it a timelike circular geodesic or a non-geodesic orbit) can
have a larger angular frequency. Thus, such spacetimes are
characterized by
\begin{equation}\label{Eq36}
\Omega_{\text{timelike}}<\Omega_{\text{null}}\
\end{equation}
for all timelike circular geodesics.

Finally, following the result (\ref{Eq35}) for the shortest possible
orbital period around a Kerr black hole as measured by asymptotic
(flat-space) observers, it is tempting to conjecture a general lower
bound on orbital periods around compact objects:
\begin{equation}\label{Eq37}
T_{\infty}\geq 4\pi M\  ,
\end{equation}
where $M$ is the mass of the central object.

\bigskip
\noindent {\bf ACKNOWLEDGMENTS}
\bigskip

This research is supported by the Meltzer Science Foundation. I
thank Yael Oren, Arbel M. Ongo and Ayelet B. Lata for stimulating
discussions.


\begin{thebibliography}{99}
\bibitem{Notephoton} The circular null geodesic is actually the limiting case
of circular timelike geodesics. That is, it is the innermost
circular orbit in black-hole spacetimes \cite{Chan,Shap,CarC}.
\bibitem{Notegeonon} It is worth emphasizing that we consider here both
geodesic and non-geodesic trajectories.
\bibitem{Notephi} Note that for $\Delta\phi=-2\pi$ the angular velocity of the
circling object and the angular velocity of the central black hole
are of opposite signs as measured by asymptotic observers.
\end{thebibliography}
\end{document}